\documentclass[journal=nalefd]{achemso}

\usepackage{achemso}
\usepackage[T1]{fontenc} 

\usepackage{amssymb}
\usepackage[table]{xcolor}

\newif\ifshowcomments\showcommentstrue      
\usepackage[final]{changes}

\newcommand{\Ir}{1T-TaIrTe$_4$}

\newcommand{\2}{$_2$}
\newcommand{\ea}{\textit{et al.}}
\newcommand{\marpes}{$\mu$-ARPES}

\author{Honey Boban}
\affiliation{Department of Quantum Matter Physics, University of Geneva, 24 quai Ernest Ansermet, CH-1211 Geneva, Switzerland}
\author{Tanguy Prongué}
\affiliation{Department of Quantum Matter Physics, University of Geneva, 24 quai Ernest Ansermet, CH-1211 Geneva, Switzerland}

\author{Amarjyoti Choudhury}
\affiliation{Dipartimento di Scienze Fisiche, Informatiche e Matematiche,  Universit\`a di Modena e Reggio Emilia, I-41125 Modena, Italy}

\author{Dario Marchiani}
\affiliation{Department of Quantum Matter Physics, University of Geneva, 24 quai Ernest Ansermet, CH-1211 Geneva, Switzerland}

\author{Salony Mandloi}
\affiliation{Department of Quantum Matter Physics, University of Geneva, 24 quai Ernest Ansermet, CH-1211 Geneva, Switzerland}

\author{Andrés Bareño}
\affiliation{Department of Quantum Matter Physics, University of Geneva, 24 quai Ernest Ansermet, CH-1211 Geneva, Switzerland}

\author{Felix Eder}
\affiliation{Department of Quantum Matter Physics, University of Geneva, 24 quai Ernest Ansermet, CH-1211 Geneva, Switzerland}
\author{Enrico Giannini}
\affiliation{Department of Quantum Matter Physics, University of Geneva, 24 quai Ernest Ansermet, CH-1211 Geneva, Switzerland}

\author{Fabian O. von Rohr}
\affiliation{Department of Quantum Matter Physics, University of Geneva, 24 quai Ernest Ansermet, CH-1211 Geneva, Switzerland}

\author{Alberto F. Morpurgo}
\affiliation{Department of Quantum Matter Physics, University of Geneva, 24 quai Ernest Ansermet, CH-1211 Geneva, Switzerland}
\alsoaffiliation{Group of Applied Physics, University of Geneva, 24 quai Ernest Ansermet, CH-1211
Geneva, Switzerland}

\author{Marco Gibertini}
\affiliation{Dipartimento di Scienze Fisiche, Informatiche e Matematiche,  Universit\`a di Modena e Reggio Emilia, I-41125 Modena, Italy}
\alsoaffiliation{Istituto Nanoscienze -- CNR, S3, I-41125 Modena, Italy}

\author{Anna Tamai}
\affiliation{Department of Quantum Matter Physics, University of Geneva, 24 quai Ernest Ansermet, CH-1211 Geneva, Switzerland}

\author{Felix Baumberger}
\email{felix.baumberger@unige.ch}
\affiliation{Department of Quantum Matter Physics, University of Geneva, 24 quai Ernest Ansermet, CH-1211 Geneva, Switzerland}
\alsoaffiliation{Swiss Light Source, Paul Scherrer Institute, CH-5232 Villigen, Switzerland}

\title{Quasi-one-dimensional topological band structure and van Hove singularities in monolayer TaIrTe$_4$ from laser $\mu$-ARPES}

\keywords{2D materials, electronic structure, quantum spin Hall insulator, nano-ARPES}

\begin{document}

\begin{abstract}
Recent transport experiments reported a quantum spin Hall insulator phase in monolayer \Ir{} gated away from charge neutrality. This phase is not predicted by band structure calculations and has been attributed to an electronic instability induced by strong correlations at a putative van Hove singularity. 
Here, we investigate the electronic structure of exfoliated monolayer \Ir{} using micro-focus laser angle resolved photoemission. 
We find a strongly anisotropic band structure susceptible to density wave instabilities.
Our data further reveal a saddle point singularity in the density of states. 
However, we find that the saddle point lies at a density for which \added{linear} transport experiments found no anomalies.
Moreover, the quasiparticle line widths suggest weak electron correlations only.
This points to a secondary role of van Hove singularities and electron correlations in the transport phase diagram of monolayer \Ir.
\end{abstract}

%

\vspace{1cm}
The interplay of correlations and topology can give rise to novel phases of matter~\cite{Dzero2010,Neupert2011,WitczakKrempa2014}. While this prospect was recognized early on, the field long remained experimentally challenging.
However, recent advances in two-dimensional (2D) moiré materials brought about remarkable progress.
The archetypal manifestation of the synergy between band topology and strong correlations is the fractional quantum anomalous Hall (FQAH) effect in Chern insulators -- 2D topological insulators with broken time-reversal symmetry. Evidence for the FQAH effect was first reported from optical measurements of twisted bilayer MoTe\2~\cite{Cai2023}. Shortly after, the direct transport observation of fractional Hall plateaus unambiguously confirmed the FQAH effect in MoTe\2 and multilayer graphene~\cite{Lu2024,Park2023}. 
Time reversal invariant quantum spin Hall insulators~\cite{Kane2005}, the first type of topological insulator predicted in 2006 and realized in 2007 in HgTe quantum wells~\cite{Bernevig2006,Koenig2007}, do not show a charge Hall effect but may also host fractional states manifest in fractional edge conduction~\cite{Bernevig2006,Levin2009,Kang2024}.
Recent studies further provide evidence for topological Kondo insulators~\cite{Dzero2010,Han2026} and anomalous Hall crystals, topologically non-trivial correlated insulators at fractional filling that spontaneously break translational symmetry and time-reversal symmetry~\cite{Dong2024}.
Topological Mott insulators have been discussed theoretically~\cite{Raghu2008,Pesin2010}, but proved elusive in experiment, in part because their signatures are unconventional and may be difficult to detect~\cite{Wagner2023}.

\Ir{} provides a promising material platform in this context. In its bulk form, \Ir{} was identified as a non-centrosymmetric type-II Weyl semimetal candidate~\cite{Koepernik2016,Haubold2017}. Upon dimensional reduction, it was predicted that monolayer and bilayer \Ir{} realize quantum spin Hall insulating phases at charge neutrality~\cite{Liu2017,Marrazzo2019,Guo2020}.
Intriguingly though, recent experiments by Tang~\ea{} revealed
two distinct insulating states with quantized edge conduction in monolayer \Ir{}.
This adds a new aspect to the rich physics of interacting topological insulators~\cite{Tang2024}.
The first insulating state, observed at charge neutrality, is consistent with density functional theory (DFT) predictions~\cite{Liu2017,Marrazzo2019,Guo2020} and is likely a single particle topological band insulator. The second insulating state is more puzzling. It was observed in gated devices at a carrier density $n_e\sim 6.5\cdot 10^{12}$~cm$^{-2}$, where the chemical potential lies a few 10 meV in the conduction band (CB), and is separated from the topological band insulator at charge neutrality by a narrow, weakly metallic region. These characteristics are incompatible with a single particle band structure picture and strongly suggest a many-body nature of the insulating gap.

Tang~\ea{} interpret the finite-density insulating state in monolayer \Ir{} as a topological charge density wave (CDW) insulator and propose that the instability is induced by enhanced electron correlations when a van Hove singularity (VHS) predicted by DFT approaches the Fermi level~\cite{Tang2024}. 
Raman, nonlinear Hall, and nonlinear optical measurements in few-layer TaIrTe$_4$ have equally been interpreted as evidence for a low-temperature 1D CDW phase~\cite{Jiang2025}.
\added{However, the putative 1D topological density wave has thus far not been established directly in experiment. Instead, very recent work found an independent bistable 2D superlattice phase, which can be induced only via cooling at high electron doping, but extends into the valence band (VB)~\cite{Tang2026}.}

\added{These experiments indicate a generic proximity of \Ir{} to density wave instabilities.}
However, several questions remain open. 
\added{Van Hove singularities have not been established from experiment, and more broadly, the electronic structure of \Ir{} on energy scales relevant for the different instabilities remains largely unknown~\cite{Tang2024,Jiang2025,Tang2026}.
}
A recent synchrotron nano-focus angle resolved photoemission (ARPES) study found a band structure in qualitative agreement with DFT on large energy scales but could not resolve the relevant low-energy band structure with the predicted VHS~\cite{Ekahana2026}.
Moreover, Tang~\ea{} find the many-body QSHI state at a much lower density than expected from the nesting vector predicted by DFT\added{, calling an interpretation of transport based on DFT band structure into question}~\cite{Tang2024,Li2026}.
Finally, a systematic study on a large number of devices revealed significant variations with some devices showing only a single QSHI phase and a third category of devices showing two topologically trivial insulating states~\cite{Li2026}.

Here, we use laser \marpes{} to map the full low-energy quasiparticle dispersion in ML \Ir. Together with DFT calculations, our data unambiguously confirm a topologically non-trivial band structure with gapped Dirac points at 4 equivalent momenta on the Brillouin zone boundary. We further demonstrate the presence of VHSs with significant density of states originating from shallow saddle points connecting the Dirac points with the global valence band maximum at $\Gamma$.

\begin{figure*}[tb]
\begin{center}
\includegraphics[width=0.7\textwidth]{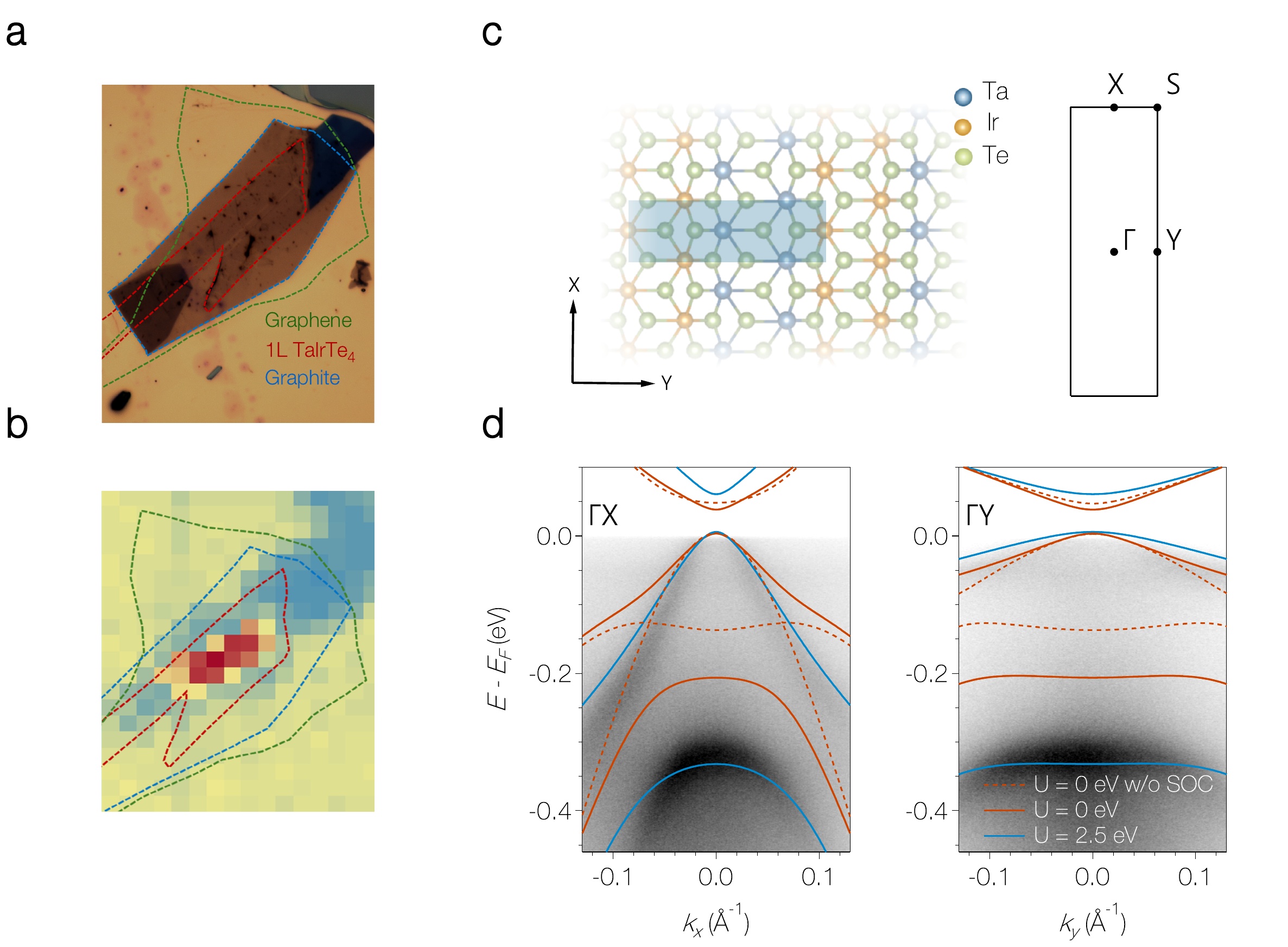}
\caption{Anisotropy of crystal structure and electronic band structure. 
(a,b) Optical micrograph and ARPES spatial mapping of an encapsulated ML \Ir{} sample measured for this study.
(c) Structure of ML \Ir{} with the primitive unit cell and lattice vectors indicated.  The Brillouin zone is shown on the right.
(d) ARPES dispersion plots along $\Gamma$X and $\Gamma$Y. DFT calculations with and without SOC and DFT+$U$ calculations (for $U=2.5$~eV) are overlaid in dashed and solid lines, respectively. All calculations are shifted in energy to match the experimental Fermi wave vector along $\Gamma$X.
}
\end{center}
\end{figure*}

Figure~1a illustrates the sample design used for this study. We exfoliate \Ir{} in a N$_2$-filled glove box. MLs are identified from the optical contrast and encapsulated between a graphite bottom layer and ML graphene using polycarbonate (PC) stamps. The stacks are placed on Au coated Si/SiO\2{} wafers and rinsed in chloroform prior to transferring the samples under protected atmosphere to the vacuum system for \marpes{}, where they are located in ARPES spatial maps (Figure~1b). All measurements were performed at $T=6$~K, a photon energy of 6.01~eV and an energy and momentum resolution of $\sim 6$~meV / $0.002$~\AA$^{-1}$.
DFT calculations in the generalized gradient approximation with on-site Hubbard corrections (GGA$+U$) were performed using the Quantum ESPRESSO package~\cite{Giannozzi2009,Giannozzi2017}.
Additional details of the experimental and theoretical methods are given in Supporting Information.

ML \Ir{} has an orthorhombic structure with alternating zigzag chains of Ta and Ir atoms in edge sharing, distorted Te octahedra (Figure~1c). 
The in-plane lattice constants measured in the bulk are $a=3.770$~\AA{} (along the chains) and $b=12.421$~\AA{} (perpendicular to the chains)~\cite{Mar1992,Xing2020}.
Importantly, the structure has a glide mirror symmetry, which protects Dirac band crossings and allows the Dirac points to move away from time-reversal invariant momenta~\cite{Muechler2016}. The product of the glide-mirror with a 2-fold screw rotation makes the structure inversion symmetric~\cite{Muechler2016,Zhang2024}.
The low-energy states of \Ir{} arise from hybridization of Ta $5d$ and Te $5p$ orbitals~\cite{Tang2024}, suggesting a strongly anisotropic electronic structure. Our \marpes{} data of ML \Ir{} shown in Figure~1d directly confirm this expectation. We observe a strong dispersion along the Ta chains ($\Gamma$X) and suppressed dispersion in the orthogonal $\Gamma$Y direction. 
Notably, our DFT calculation shown in the same figure underestimates the quasiparticle dispersion along $\Gamma$X by nearly a factor of 2, opposite to the expectations for a correlated material with enhanced quasiparticle mass (see Figure~1d).
Investigating the origin of this discrepancy, we notice that \Ir{} has a light and heavy valence band, found in our ARPES data at $E_F$ and $E-E_F\approx 0.3$~eV, respectively.
\added{In DFT} without SOC, these bands cross at 2 degeneracy points along $\Gamma$X at $E-E_F \sim -0.13$~eV (Figure~1d). Adding SOC lifts the degeneracies by hybridizing the two bands.
This renders their dispersion highly sensitive to their relative energy position.
Here, we use DFT+$U$ as a pragmatic way to tune the latter. 
Empirically, we find that adding $U\approx 2.5$~eV on Ta and Ir sites correctly reproduces the experimental valence band positions at $\Gamma$ and at the same time improves the overall description of the experimental quasiparticle dispersion.
For the remainder of this paper, we will refer to this DFT+$U$ calculation. The systematic evolution of the band structure with $U$ is shown in Supporting Information, Figure~S1.

\begin{figure*}[tb]
\begin{center}
\includegraphics[width=0.95\textwidth]{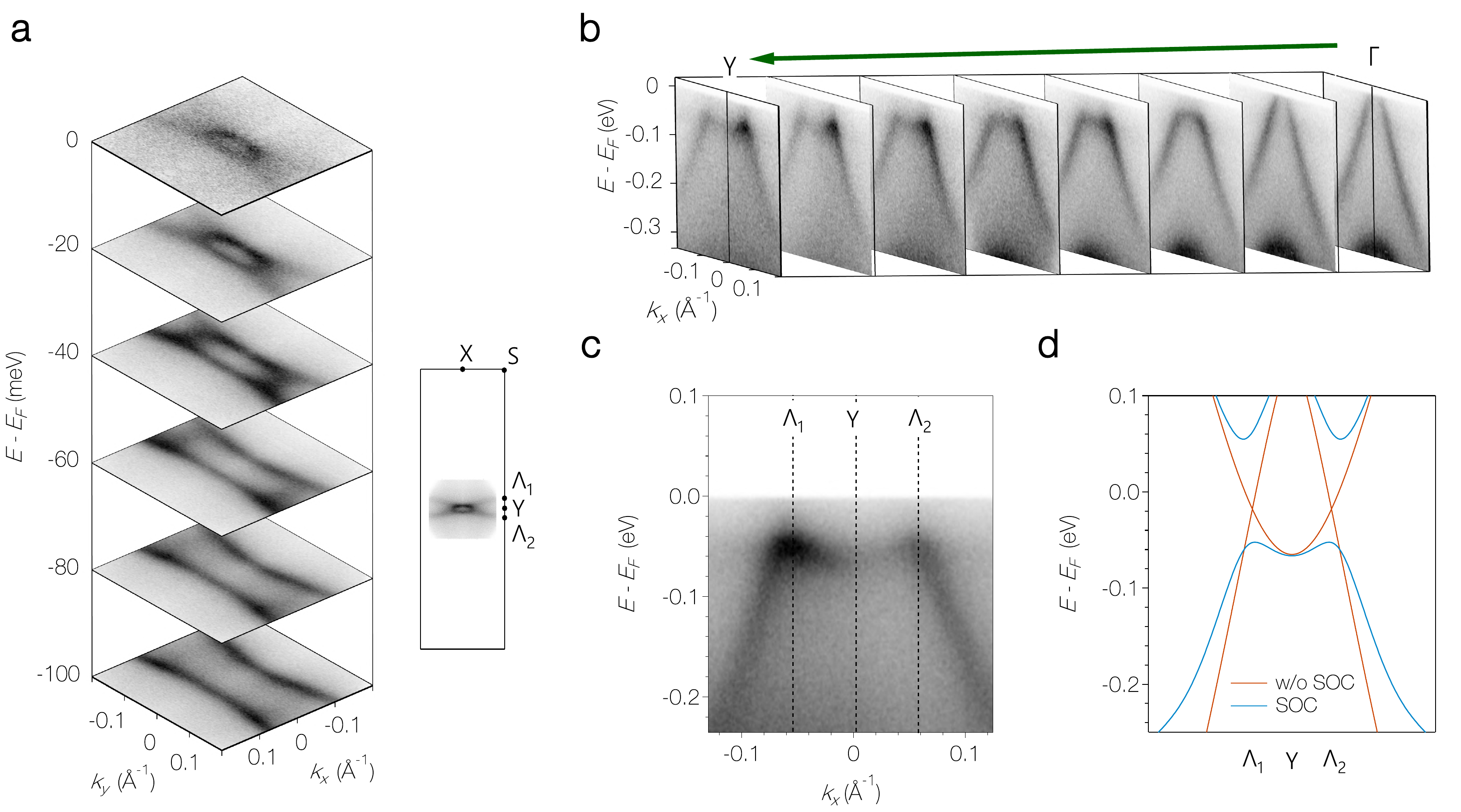}
\caption{Gapped Dirac cones and low-energy quasiparticle dispersion. (a) Stack of constant energy contours. For scale, the spectral weight at $E_F$ is also shown in the full Brillouin zone.
(b) Dispersion plots along $k_x$ for equidistant $k_y$ values ranging from $\Gamma$ to Y.  (c) Dispersion plot along the edge of the BZ. (d) DFT+$U$ band dispersion showing the gapping of the Dirac cones with the inclusion of SOC. 
}
\end{center}
\end{figure*}

In Figure~2, we focus on the low-energy quasiparticle band structure. 
Consistent with Figure~1, we find that our samples are degenerately doped with the chemical potential lying $\sim 7$~meV below the valence band maximum. 
The Fermi surface (top panel in Figure~2a, also shown to scale in the full Brillouin zone) consists of a very small elliptical hole pocket at $\Gamma$ 
with a carrier density of $7\cdot 10^{11}$~cm$^{-2}$ (0.0016 holes per Ta). At present, it is not clear if the hole doping arises from defects in our crystals or is induced by charge transfer to the graphite and graphene encapsulation layers. 
With increasing energy, the ARPES intensity first evolves into an unusual hashtag like structure before developing the characteristic warped lines of a quasi-1D band structure. The series of dispersion plots in Figure~2b reveals that the peculiar hashtag contours arise from the evolution of the $\Gamma$ valley into an extended $k$-space area with nearly flat states at an energy $\sim -60$~meV. 

Figure~2c shows an expanded view of the band map along the Brillouin zone boundary. The abrupt change in dispersion near the $\Lambda$ points is a fingerprint of an avoided crossing between a light hole band and a heavier electron-like band. Our DFT+$U$ calculations (Figure~2d) readily identify these bands as Ta $d$ states with two symmetry-protected Dirac crossings along SYS. The inclusion of SOC lifts the degeneracy, resulting in topological band gaps at $\Lambda_{1,2}$, consistent with earlier DFT studies~\cite{Liu2017,Marrazzo2019,Guo2020}. Adding correlations in DFT+$U$ affects band gaps and reshapes the band structure. 
The gap at $\Lambda$ increases from 69~meV at $U=0$ to 108~meV at $U=2.5$~eV and moves to Y for $U\geq 4$~eV (see Supporting Information, Figure~S1). 
Importantly though, we find that the quantum spin Hall state of \Ir{} is robust in DFT+$U$ up to at least $U=5$~eV.
The valence band maximum lies at $\Gamma$ in all our calculations. At $U=2.5$~eV, the valence band valley splitting $E_{\Gamma}-E_{\Lambda}$ is 59~meV, close to the experimental value of $\approx 57$~meV obtained from $E_{\Lambda}=-50$~meV (Figure~2c) and $E_{\Gamma}\approx 7$~meV estimated in Figure~1.
The local conduction band minima at $\Gamma$ and $\Lambda$ are not detected in our experiments. The calculations show them nearly degenerate with the direct and indirect gap both increasing slowly with $U$ from 33~meV at $U=0$ to 49~meV at $U=2.5$~eV, slightly larger than the gaps of $5 - 30$~meV deduced from transport experiments~\cite{Tang2024,Li2026}.

\begin{figure*}[tb]
\begin{center}
\includegraphics[width=0.9\textwidth]{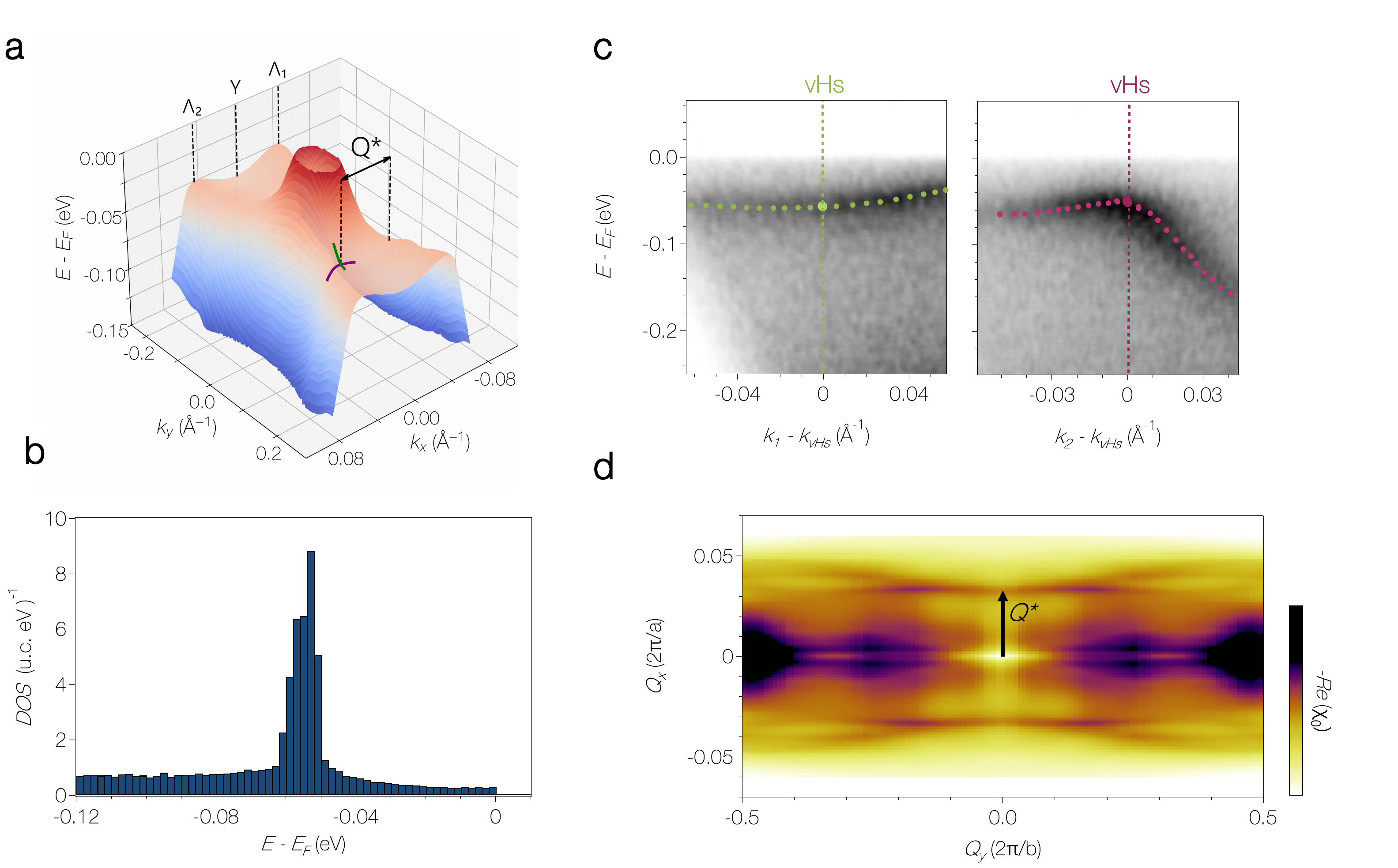}
\caption{Van Hove singularity. (a) ARPES quasiparticle dispersion obtained from the analysis of EDCs and MDCs.
(b) Experimental quasiparticle DOS. 
(c) Dispersion along the two $k$-space paths intersecting at the VHS indicated in panel~a. (d) Static Lindhard susceptibility calculated from the experimental quasiparticle dispersion.
}
\end{center}
\end{figure*}

The band structure of \Ir{} has two key properties that favor many-body instabilities. It is quasi-1D and in addition shows extended flat regions. In Figure~3 we analyze these properties quantitatively. To this end, we first determine in Figure~3a the full 2D low-energy quasiparticle band structure $E(k_x,k_y)$ from an analysis of energy- and momentum distribution curves (EDCs, MDCs). 
Figure~3b shows the quasiparticle density of states (DOS) obtained from a histogram of the $E(k_x,k_y)$ values. The pronounced peak at $-52$~meV can be attributed to a VHS at a filling of $6.2\cdot10^{12}$~cm$^{-2}$.
We directly confirm the presence of VHSs in Figure~3c. Dispersion plots extracted along the intersecting $k$-space paths marked green and violet in Figure~3a show a shallow saddle point with opposite curvature along $k_{1,2}$, implying a VHS with diverging DOS. At the same time, it is evident from Figure~3c that a minute deformation of the bands could transform the saddle into a ridge and move the VHS out to the $\Lambda_{1,2}$ points, or could even merge two singularities at the Y point, as we observe it in DFT+$U$ for $U\geq 4$~eV (see Supporting Information, Figure~S1).
Supporting Figure~S2 further compares the experimental quasiparticle DOS with DFT and DFT+$U$.


\added{Intriguingly, the VHS filling mirrors the CB density of the many-body QSHI phase~\cite{Tang2024}, and coincides with the density at which the non-linear Hall signal characteristic of the bistable 2D superlattice state is observed~\cite{Tang2026}. However, this does not imply a causal relationship. We first note that the filling of a CDW insulator with 1D superlattice vector $\bf{Q^*}$ connecting VHSs differs from the VHS filling. In the specific case of ML \Ir, this difference is substantial.}
Figure~3d shows the static susceptibility $\chi^0({\bf q})$, calculated from the experimental quasiparticle dispersion, assuming a rigid band shift of 52~meV to move the VHS to the Fermi level.
\added{This shows multiple quasi-1D features in a narrow range of $Q_a\lesssim 0.04\:(2\pi/a)$. Notably, at the periodicity $\sim 9.3a \times 14.2 b$ of the bistable superlattice state identified by STM (${\bf Q}\sim(0.11\frac{2\pi}{a}, 0.07\:\frac{2\pi}{b})$)~\cite{Tang2026}, we find a negligible electronic susceptibility only.}
The most prominent feature at ${\bf Q_{\mathrm{ARPES}}^*}=0.031\:(2\pi/a)$ corresponds to nesting between two VHSs and closely resembles the structure in the bare DFT susceptibility at ${\bf Q_{\mathrm{DFT}}^*}=0.052\:(2\pi/a)$ identified by Tang~\ea{} in Ref.~\cite{Tang2024} as origin of the putative CDW instability. 
\added{However, filling the first band in a 1D superlattice with ${\bf Q_{\mathrm{ARPES}}^*}=0.031\:(2\pi/a)$, requires a density $n= Q_{\mathrm{ARPES}}^* a/\pi=0.062/u.c.$ ($1.3\cdot 10^{13}$~cm$^{-2}$), roughly a factor of 2 higher than the VHS filling and beyond the range investigated by Tang~\ea~\cite{Tang2024}.}
\added{In addition, the CB structure may differ from the VB probed here.}
Consistent with Ref.~\cite{Tang2024}, our DFT calculations for $U=0$ show a nearly particle-hole symmetric band structure with a single singularity in both  the conduction and valence band DOS. However, for $U=2.5$~eV, we find a splitting of more than 20~meV in the conduction band singularity, while the valence band singularity stays intact (Supporting Information, Figure~S2). It remains to be understood if this qualitative change underlies the different behavior of \Ir{} under electron and hole doping observed in experiment~\cite{Tang2024,Tang2026}.

\added{The transport signatures near VHS filling are equally unusual.}
If a system is tuned to VHS filling, one expects anomalous transport from scattering with 'hot' van Hove carriers leading to an increased resistivity and deviations from the canonical $T^2$ behavior~\cite{Mousatov2020}, as was observed in moiré systems or in Sr$_2$RuO$_4$~\cite{Wei2024,Barber2018}. However, thus far, such anomalies have not been reported in \Ir{} gated to van Hove filling. 
\added{Instead, our data show that the strong non-linear Hall signal observed in the bistable superlattice phase occurs at - or near - VHS filling~\cite{Tang2026}.} 

\begin{figure*}[tb]
\begin{center}
\includegraphics[width=0.39\textwidth]{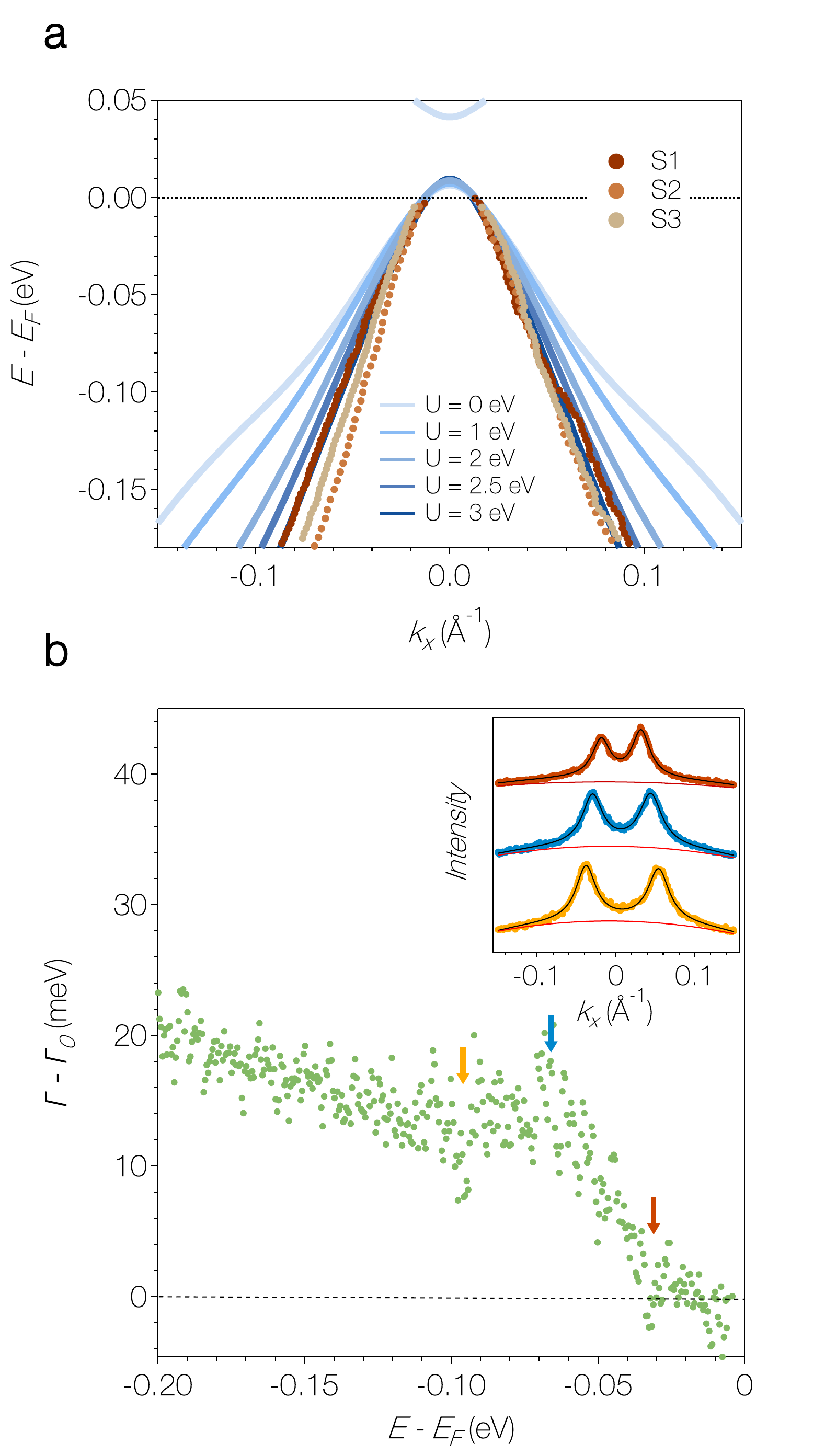}
\caption{Many-body interactions. (a) Comparison of the quasiparticle dispersion along $\Gamma$X from three devices (S1, S2, S3) with the DFT+$U$ bands. (b) Quasiparticle linewidth $\Gamma=\partial E/\partial k\:\Delta k$ calculated 
\added{from the full width at half maximum $\Delta k$ of Lorentzian MDC fits and a smooth dispersion $\partial E/\partial k$ obtained from a polynomial fit of the MDC peak positions. An energy independent contribution from defect scattering has been subtracted.} The inset shows fits of MDCs at representative energies marked by vertical arrows of the same color. Red lines are the background of the fit.
}
\end{center}
\end{figure*}

This all points to a secondary role of VHSs in \Ir. Instead, the properties of ML \Ir{} may be dominated by 
\replaced{its peculiar topological electronic structure revealed here.}
{the unusual electronic structure of \Ir{} with extended regions with nearly flat bands.}
We first note that the periodicity of CDWs is often determined by a strongly momentum dependent electron-phonon interaction and rarely coincides with nesting vectors~\cite{Johannes2008}.
\added{The topological band inversion in ML \Ir{} creates extended regions around the VHSs where the CB is nearly flat.}
Placing the chemical potential near or inside this flat band naturally leads to a large susceptibility at small ${\bf Q}$. Calculations using the experimental quasiparticle dispersion show that such small-${\bf Q}$ features may exceed the susceptibility peak from nesting of the VHS (Supporting Information, Figure~S3). In this situation, the precise wave vector of an instability is likely determined by the momentum dependence of electron-phonon coupling. \added{At the same time, the topological band inversion naturally leads to a large Berry curvature in the flat band, which may explain the non-linear Hall effect.
}

The combination of flat bands and substantial Coulomb repulsion in the Ta and Ir d-shell suggest a potentially important role of electron correlations in \Ir.
The strength of correlations is often assessed from the renormalization of the quasiparticle Fermi velocity over the bare dispersion. However, this requires an accurate bare Hamiltonian -- typically obtained from DFT in the local density approximation~\cite{Tamai2019}. We presently cannot give such a Hamiltonian for ML \Ir.
The primary effect of the Hubbard correction $U$ in our DFT+$U$ calculations is to tune the relative energy position of bands, and with it their hybridization. 
However, for $U=2.5$~eV, which reproduces the relative energy position of the valence bands at $\Gamma$, our calculations still underestimate the quasiparticle dispersion (Figure~4a).

The evolution of the quasiparticle linewidth provides an alternative, albeit less quantitative, measure of interactions.
In strongly correlated materials, the linewidth typically exceeds the excitation energy beyond a few 10 meV from the Fermi level where defined quasiparticles gradually evolve into incoherent excitations~\cite{Tamai2008,Mravlje2011}.
Our dispersion maps show no sign of such a crossover.
For a quantitative analysis, we use Lorentzian fits of MDCs along $\Gamma$X and calculate quasiparticle linewidths $\Gamma$ \added{using a smooth group velocity obtained from a polynomial fit of MDC peak positions.} This reveals an exceptionally slow variation with energy. As shown in Figure~4b, $\Gamma$ increases  \replaced{barely 20~meV}{barely more than 30~meV} over the full energy range of $\sim 200$~meV. This differs markedly from typical correlated electron systems~\cite{Koralek2006,Smit2024} and is more comparable to nearly free electron metals, pointing to weak correlations~\cite{Potorochin2022}.
We note, however, that correlation effects can be highly sensitive to the band filling. Intriguingly, our analysis in Figure~4b shows a small but statistically significant linewidth anomaly at the energy of the VHS, pointing to possibly relevant electron correlations around this filling.

In summary, we reported a comprehensive mapping of the low-energy quasiparticle dispersion in ML \Ir. 
Our data provide strong evidence for a topological band inversion and directly resolve low-lying van Hove singularities. 
This provides a solid foundation for the interpretation of complementary transport experiments.
Our findings broadly support the notion of a topological CDW insulator in gated ML \Ir{} put forward by Tang~\ea~\cite{Tang2024}.
However, they also reveal a mismatch between the density of the many-body QSHI state in ML \Ir{} and the wave vector of the most prominent feature in the single particle susceptibility. 
This highlights the need for structural studies and for future ARPES experiments on gate-tuned devices to probe the finite-density insulating regime directly and search for spectral signatures of density wave order and enhanced many-body effects.

\section{Supporting Information}
\added{Additional details on experimental and theoretical methods, additional data and density functional theory calculations.
}

\section{Acknowledgments}
We thank I. Gutiérrez-Lezama for help with the development of the device fabrication.
The experimental work was supported by the Swiss National Science Foundation (SNSF) through grants 200020-215548 and 200021-204065. 
A.C.\ and M.G.\ acknowledge financial support from the EMPEROR project, CUP E93C24001040001, funded
by the European Union—NextGeneration EU (M4C2INV1.3) through the National Quantum Science and Technology Institute (Spoke 5).

\newpage

\begin{suppinfo}
\setcounter{figure}{0}
\renewcommand{\thefigure}{S\arabic{figure}}

\section{Crystal growth}
Single crystals of \Ir{} were grown using the self-flux technique from the corresponding elements. Powders of Ta (99.95 \%, Thermo Scientific), Ir (99.95 \%, Alfa Aesar) and Te (ground from lumps, 99.999 \%, Thermo Scientific) were mixed together in molar ratios of 1:1:30 inside an alumina crucible. This was placed inside an open quartz ampoule, and a ceramic sieve followed by 2--3 cm of compressed quartz wool were placed on top. Then, the ampoule was sealed off under an inert atmosphere of 300 mbar Ar and placed vertically in a muffle furnace. The temperature program consisted of heating to $1000 ^{\circ}$C with a ramp of $+30 ^{\circ}$C/h, holding the maximum temperature for 48 h before cooling down with $-2 ^{\circ}$C/h to $600 ^{\circ}$C. After a few hours at this temperature, the ampoule was taken out of the furnace, flipped around, and excess Te-flux was removed through hot-centrifugation at $ca.$ 2000 rpm. Needle-shaped crystals of the title compounds with a length of several mm could be isolated.

The elemental composition of the grown \Ir{} crystals was confirmed using energy dispersive X-ray spectroscopy (EDS) on a JEOL JSM-7600F scanning electron microscope, equipped with an X-Max\textsuperscript{N} 80 detector (Oxford Instruments), using an accelerating voltage of 20~kV. In total, 90 point-analyses were performed on 14 crystals originating from six batches. On average, a composition of Ta\textsc{1.00(3)}Ir\textsc{1.00(3)}Te\textsc{3.95(9)} was obtained, which lies within one standard deviation from the nominal composition. The crystal structure of \Ir{} was confirmed with both single-crystal (on a \textit{Oxford Diffraction SuperNova} diffractometer from Agilent/Rigaku) and powder (\textit{Rigaku SmartLab}) X-ray diffraction.

\section{Sample preparation}
Mechanical exfoliation using scotch tape and polydimethylsiloxane (PDMS) was employed to obtain monolayers of \Ir{} from the bulk crystals. Single crystals of TaIrTe$_4$ were first placed on scotch tape and exfoliated 3 to 4 times to obtain thin flakes, which were then transferred onto the PDMS. The PDMS with flakes was subsequently brought into contact with a 290 nm SiO$_2$/Si substrate and heated to $100 ^{\circ}$C for 1 minute. After heating, the PDMS was slowly peeled off from the SiO$_2$ substrate at room temperature, transferring the flakes on the substrate.\\
The thickness of TaIrTe$_4$ flakes on the SiO$_2$/Si substrate was initially estimated using optical contrast analysis. Monolayer TaIrTe$_4$ exhibited normalized contrast values of approximately 7.5\%, 11\%, and 2.5\% in the red, green, and blue channels, respectively. These thickness assignments were later confirmed by ARPES measurements.\\
For ARPES measurements, the monolayer TaIrTe$_4$ flake was encapsulated in between monolayer graphene and bulk graphite. The top graphene layer serves as a capping layer, while the bottom graphite flake provides a flat and conductive substrate. The heterostructure was assembled using a PDMS/polycarbonate (PC) stamp by dry transfer technique. In this process, the topmost flake (graphene) was first picked up, followed by the monolayer TaIrTe$_4$ and the graphite flake. During pickup, the substrate temperature was initially set to $90^{\circ}$C upon contact of the stamp and ramped to $120^{\circ}$C before cooling down. After the pick up, the heterostructure along with the PC film, was released on a gold -coated SiO$_2$/Si substrate at $180^{\circ}$C. Finally, the sample was immersed in chloroform overnight to dissolve the PC layer and to remove residual polymer contamination. \\
The fabricated sample was transferred to the load-lock chamber of the ARPES system under a protected environment to prevent contamination and oxidation from ambient exposure. All processing steps, including exfoliation, heterostructure assembly, and sample mounting, were carried out inside a nitrogen-filled glovebox with impurity levels maintained below 0.1 ppm for both H$_2$O and O$_2$.

\begin{figure*}[tb]
\begin{center}
\includegraphics[width=0.98\textwidth]{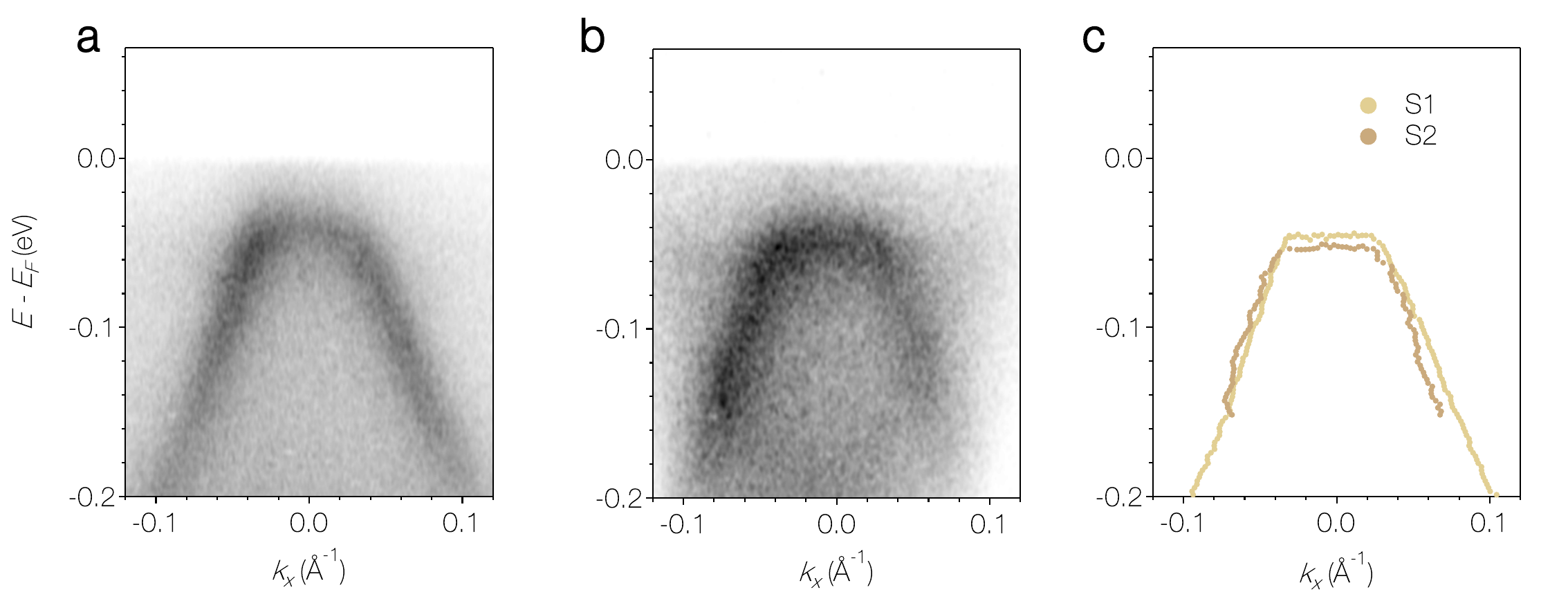}
\caption{Dispersion data along $k_x$ for $k_y= 0.09$~\AA$^{-1}$. (a,b) Raw data for devices S1 and S2 from Figure~4 of the main text. (c) Comparison of the quasiparticle dispersion extracted from the data.}
\label{sfig:ARPES}
\end{center}
\end{figure*}

\section{Reproducibility}
Fig.~4 of the main text shows good reproducibility of the data along the $\Gamma$X high-symmetry line $(k_x)$. In Figure~\ref{sfig:ARPES}, we additionally compare the dispersion of two devices along $k_x$ for $k_y= 0.09$~\AA$^{-1}$, roughly $2/3$ out towards the VHS. This shows a device to device variation of the flat band position of $\sim 5$~meV. Together with the good reproducibility of the overall dispersion, this suggests a similar uncertainty of the VHS position, which is small compared to the VHS energy of $-52$~meV.

\begin{figure*}[tb]
\begin{center}
\includegraphics[width=0.98\textwidth]{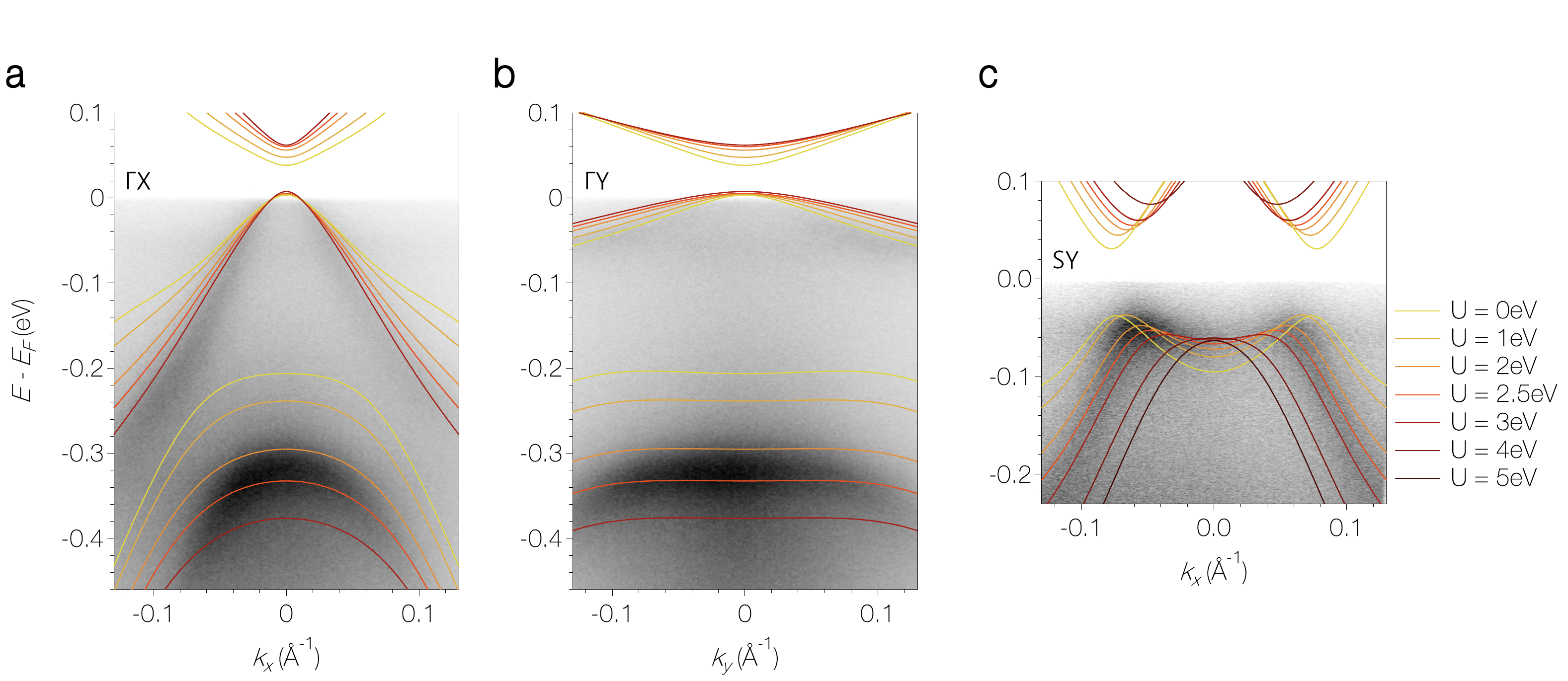}
\caption{DFT+$U$ calculations showing a systematic evolution of the electronic band structure with on-site Coulomb correction $U$. Fermi levels are chosen to reproduce the experimental Fermi wave vectors along $\Gamma$X. The ARPES data reproduced from Figures~1,2 is shown for comparison.}
\label{sfig:DFT_U}
\end{center}
\end{figure*}

\section{DFT calculations}
Density functional theory (DFT) calculations were carried out using the Quantum ESPRESSO package~\cite{Giannozzi2009,Giannozzi2017}, employing the PBE~\cite{PBE1996} exchange$-$correlation functional together with pseudopotentials obtained from the PseudoDojo~\cite{Hamann2013,PseudoDojo2018} library. The kinetic energy cutoffs for the wavefunctions and charge density were chosen as 90 Ry and 360 Ry, respectively, for the DFT simulations of the \Ir{} two-dimensional monolayer. A $\Gamma$-centered Monkhorst$-$Pack (12 $\times$ 4 $\times$ 1) k-point grid was used to sample the Brillouin zone (BZ). A cold smearing of 0.02 Ry was employed, with the input files generated using the Materials Cloud Quantum ESPRESSO input generator. To eliminate spurious electrostatic interactions between periodically repeated images along the out-of-plane direction, a Coulomb cutoff~\cite{Rozzi2006,Sohier2017} scheme was employed, ensuring appropriate boundary conditions.  The electronic band structure was calculated using the experimental crystal structure, with spin--orbit coupling incorporated through fully relativistic pseudopotentials. To achieve better agreement between the experimental and calculated electronic band structures, the GGA+$U$ approach was employed by applying on-site Hubbard corrections to the Ir-$d$ and Ta-$d$ orbitals. Several combinations of $U$ values were examined, including $(0,0)$, $(0,1)$, $(1,0)$, $(0,2)$, $(2,0)$, $(1,2)$, $(2,1)$, $(2,2)$, and $(2.5,2.5)$ eV for the Ta and Ir atoms, respectively. 
These show that adding $U$ on the Ta atoms dominates the effects on the low-energy band structure. For simplicity, we thus focus on equal $U$ values on Ta and Ir. Figure~\ref{sfig:DFT_U} shows the systematic evolution of the band structure with $U$. The best agreement with experiment is found for $U=2.5$~eV on both Ir-$d$ and Ta-$d$ orbitals.

In Figure~\ref{sfig:DFT_dos} we further compare the experimental quasiparticle DOS with DFT+$U$ for $U=0$ and $U=2.5$~eV.

\begin{figure*}[tb]
\begin{center}
\includegraphics[width=0.6\textwidth]{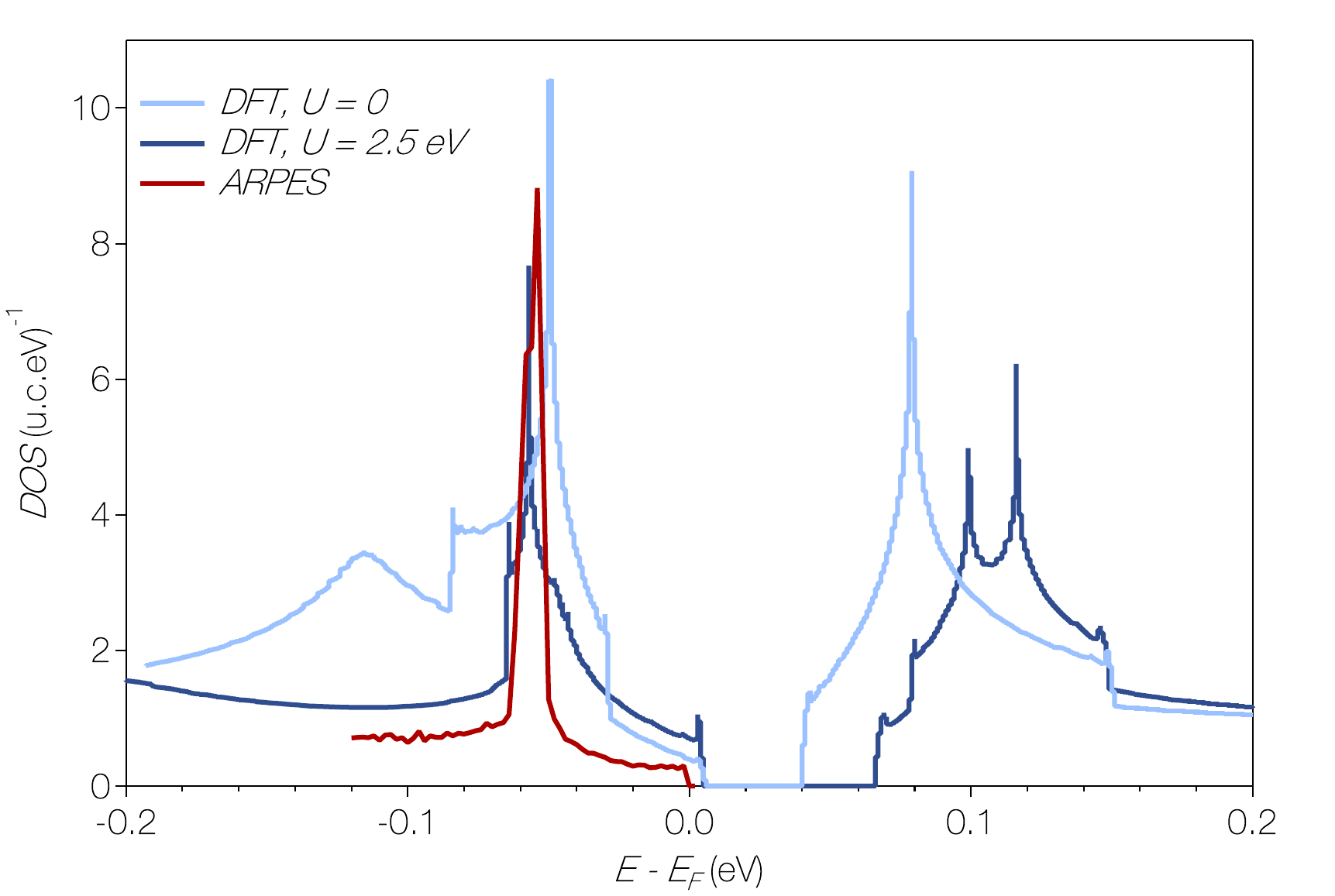}
\caption{DFT+$U$ DOS for $U=0$ and $U=2.5$~eV. Fermi levels are chosen to match the slight hole doping observed in experiment. The experimental quasiparticle DOS from Figure~3 is shown for comparison.}
\label{sfig:DFT_dos}
\end{center}
\end{figure*}

\section{Susceptibility calculation}
The charge susceptibility for monolayer \Ir{} shown in Figure~3d of the main text was computed from the static Lindhard function
\begin{equation}
\chi_0(\mathbf{Q}, \omega = 0, T) = \frac{1}{\Omega} \sum_{\mathbf{k}} \frac{f(E_{\mathbf{k}}) - f(E_{\mathbf{k}+\mathbf{Q}})}{\omega + E_{\mathbf{k}} - E_{ \mathbf{k}+\mathbf{Q}} + i\delta},
\end{equation}
using the experimental band dispersion $E_{\mathbf{k}}$ shown in Figure~3a.
Here, $\Omega$ is the system volume, $f(E_{\mathbf{k}})$ is the Fermi function and $\delta$ is a broadening parameter set to 5~meV. The temperature was fixed to $T=4.2$~K. 
The calculation in Figure~3a is performed for a rigid band shift by $\mu=-52$~meV to place the VHSs at the chemical potential.
Figure~\ref{sfig:chi_profiles} shows the systematic evolution of $\xi_0$ with band filling. The susceptibility is strongly enhanced near van Hove filling where it develops a well defined peak from nesting of the VHSs around $k_x = 0.05$~\AA$^{-1}$. In addition, the calculations show a broad local maxima at lower momentum if the chemical potential is placed within the nearly flat region of the valence band around the van Hove filling.

\begin{figure*}[tb]
\begin{center}
\includegraphics[width=0.6\textwidth]{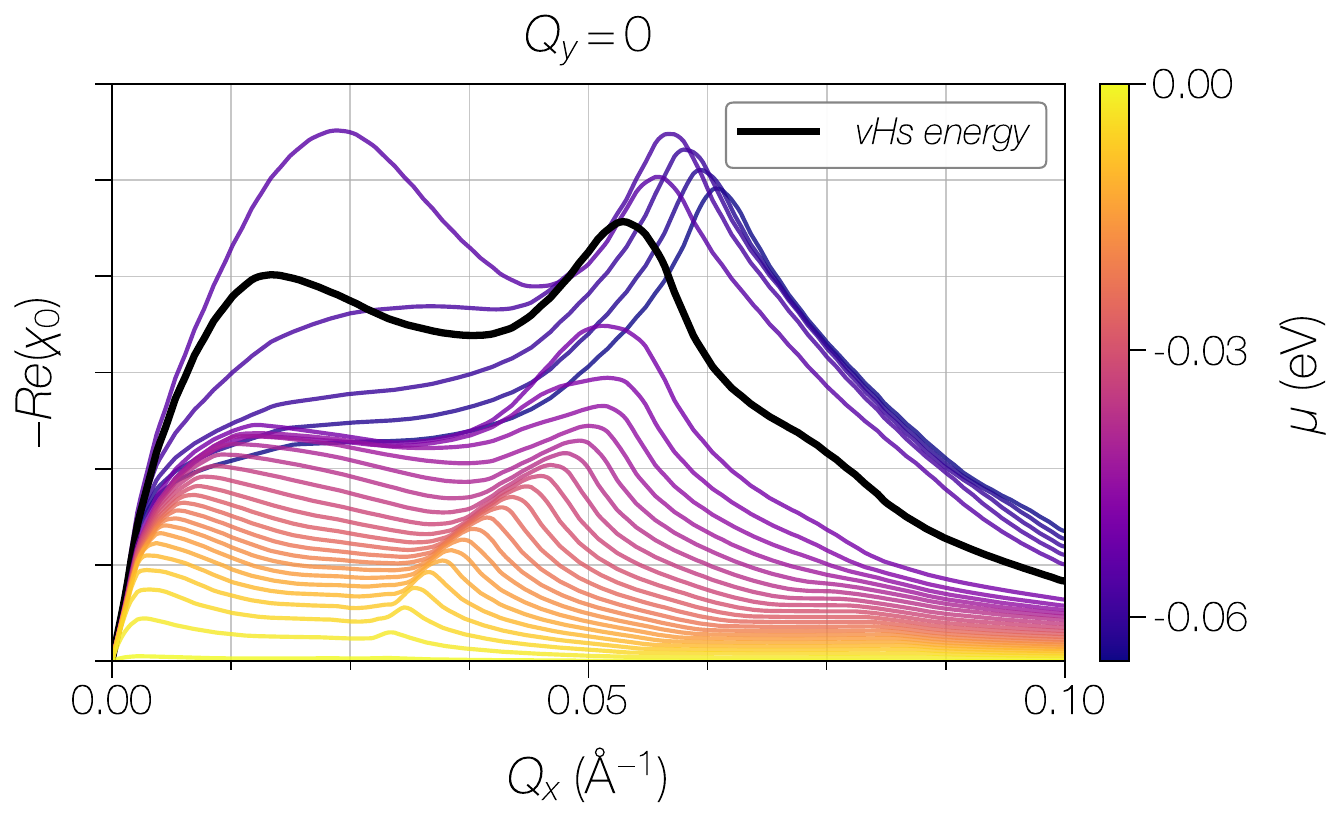}
\caption{Line profiles of the charge susceptibility $\chi_0(Q_x,Q_y =0)$ for different band fillings.}
\label{sfig:chi_profiles}
\end{center}
\end{figure*}

\end{suppinfo}

 \newpage
 \begin{figure*}[p]
 \centering
 \includegraphics[width=0.7\textwidth]{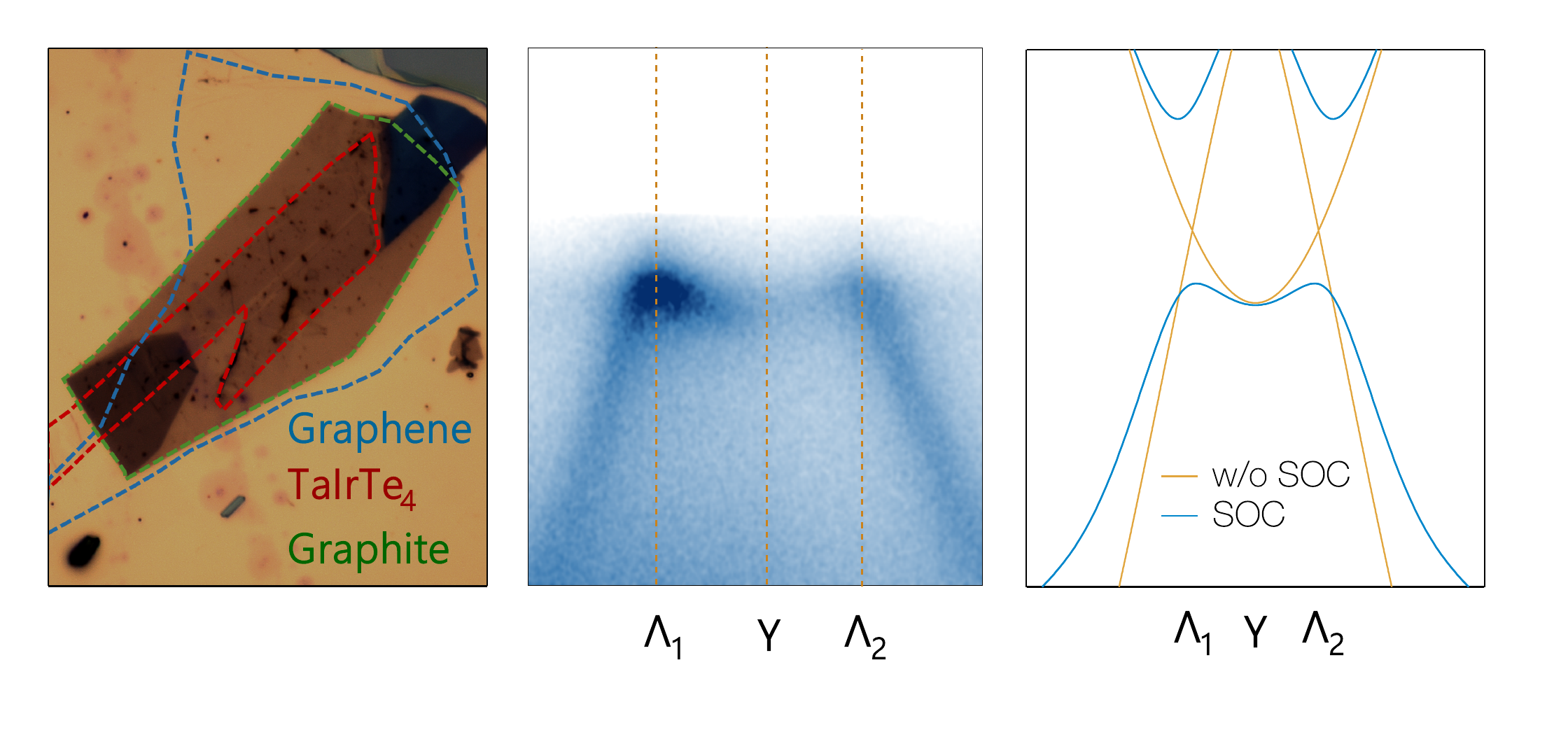}
 \caption{TOC graphic.}
 \end{figure*}
 \newpage

\providecommand{\latin}[1]{#1}
\makeatletter
\providecommand{\doi}
  {\begingroup\let\do\@makeother\dospecials
  \catcode`\{=1 \catcode`\}=2 \doi@aux}
\providecommand{\doi@aux}[1]{\endgroup\texttt{#1}}
\makeatother
\providecommand*\mcitethebibliography{\thebibliography}
\csname @ifundefined\endcsname{endmcitethebibliography}
  {\let\endmcitethebibliography\endthebibliography}{}

\end{document}